# Direct observation of orbital driven strong interlayer coupling in puckered two-dimensional $PdSe_2$

Jung Hyun Ryu[1†], Jeong-Gyu Kim[2†], Bongjae Kim[1†], Kyoo Kim[3], Sooran Kim[4], Byeong-Gyu Park[5], Younghak Kim[5], Kyung-Tae Ko[6]*, and Kimoon Lee[1]*

[1]Department of Physics, Kunsan National University, Gunsan 54150, Republic of Korea.

[2]Max Planck POSTECH/Hsinchu Center for Complex Phase Materials and Department of Physics, Pohang University of Science and Technology (POSTECH), Pohang 37673, Republic of Korea.

[3]Korea Atomic Energy Research Institute (KAERI), Daejeon 34057, Republic of Korea.

[4]Department of Physics Education, Kyoungpook National University, Daegu 41566, Republic of Korea.

[5]Pohang Accelerator Laboratory, Pohang University of Science and Technology (POSTECH), Pohang 37673, Republic of Korea.

[6]Korea Basic Science Institute (KBSI), Daejeon 34133, Republic of Korea.

*e-mail : kkt0706@kbsi.re.kr, kimoon.lee@kunsan.ac.kr

[†]These authors are equally contributed.

**Abstract**

Interlayer coupling between individual unit layers has played a critical role for layer-dependent properties in two-dimensional (2D) materials. While recent studies have revealed the significant degrees of interlayer interactions, the overall electronic structure of the 2D material has been mostly addressed by the intralayer interactions. Here, we report the direct observation of a highly dispersive single electronic band along the interlayer direction in puckered 2D $PdSe_2$ as an experimental hallmark of strong interlayer couplings. Remarkably large band dispersion along $k_z$-direction near Fermi level, which is even wider than the in-plane one, is observed by the angle-resolved photoemission spectroscopy measurement. Employing the X-ray absorption spectroscopy and density functional theory calculations, we reveal that the strong interlayer coupling in 2D $PdSe_2$ originates from the unique directional bonding of Pd *d* orbitals associated with unexpected Pd $4d^9$ configuration, which consequently gives rise to the strong layer-dependency of the band gap.

## Introduction

Since the discovery of graphene, diverse series of two-dimensional (2D) systems have been constantly emerged up to date mainly by exploring material groups with a layered structure[1-10]. These layered systems have attracted enthusiastic attention not only from fundamental physics but also from the wide field of material science because of the unique electronic structure and the additional scalability that comes from their anisotropic bonding properties. By utilizing the anisotropic character, decoupling the intralayer bonding from the van der Waals (vdW) one becomes a key approach of the 2D material science. One of the representative examples is the 2D electronic confinement down to an atomic scale that has been experimentally reached employing simple physical and/or chemical exfoliation techniques[1-10]. Such an approach has led to the various exotic low dimensional physics such as non-trivial topology[2,3], valleytronics[2,4], exotic magnetism[5,6], and superconductivity[2,7,8]. On the other hand, for the practical application utilizing 2D systems, layer-dependent properties are one of the critical issues as an effective functionality tuning strategy[2,8-16]. One of the most prototypical examples is the band gap ($E_g$) modification by controlling the number of layers[9-17]. While it depends on the class of materials, the $E_g$ widening with switching between indirect and direct gap nature has been generally observed upon decreasing the number of layers, as being a central issue for the opto-electronic applications of 2D materials[9-17]. The interlayer coupling has been addressed as the main origin for such layer-dependent properties in 2D materials[2,9-18], but it is still unclear why such layer-dependency can arise from the weak vdW-type systems.

Interlayer vdW interactions in 2D materials are considered to be weak and to have a negligible electronic contribution compared to the intralayer interactions. This is because each layer is charge-neutral and the source of interaction is dipole-dipole type, which varies in the order of $1/r^6$, hence giving larger interlayer atomic distances compared to intralayer ones[19]. However, recent studies suggest sizable electronic contributions exist despite the large interlayer space, especially for the 2D systems with strong layer-dependency[9-18]. Puckered 2D $PdSe_2$ provides an intriguing example of such thickness dependent properties in 2D materials[12-17]. On top of exceptionally low-symmetry, as similar to black phosphorus (BP), the unusual *d*-orbital occupation associated with the unique local coordination offers another degree of freedom[12,13,20]. Indeed, it is reported that 2D $PdSe_2$ is inherent with highly tunable $E_g$ and carrier polarity upon changing the number of 2D layers[12-17], possibly originating from the unusual interlayer couplings[16]. Since Oyedele *et al.* have firstly reported the wide $E_g$ variation as comparable that from BP[13], a unique thickness-dependent nonlinear optical activity has also been observed as implying the exceptional interlayer coupling behavior in 2D $PdSe_2$[17]. However, the physical origins of such strong interlayer coupling and related electronic structures have not been fully explored yet.

Here, we present the direct evidence of the exceptionally strong interlayer coupling in puckered 2D $PdSe_2$. Angle-resolved photoemission spectroscopy (ARPES) measurements on the 3D momentum space of $PdSe_2$ single crystal clearly reveal the highly dispersive electronic band along the interlayer direction, which is even more prominent than the intralayer bands. Accompanied with X-ray absorption spectroscopy linear dichroism (XLD) measurements and detailed DFT studies, we show the exceptional orbital occupation and charge character, which is due to the formation of Se-Se dimer and the crystal field splitting in a square planar symmetry, as the origin of the highly dispersive Γ-Z bands. We provide a complete description of the strong interlayer coupling and give a physical/chemical explanation on the layer number dependence of $E_g$.

## Results and discussion

Figure 1a shows the X-ray diffraction (XRD) pattern for the cleaved plane of the $PdSe_2$ sample. (The inset of Figure 1a) Collective reflection peaks solely associated with (00*l*) reciprocal momenta indicate that the cleaved surface is well-oriented parallel to the *ab*-plane direction as illustrated in the schematic crystalline structure (See the inset schematic of Figure 1b)[13-15]. The measured Raman spectrum of the cleaved single crystalline $PdSe_2$ shows four representative Raman active modes. (Figure 1b) These modes mainly involve the vibrations of the Se atoms, assigned as $A_g^1$-$B_{1g}^1$, $A_g^2$, $B_{1g}^2$, and $A_g^3$, which are consistent with previous reports[13-16]. X-ray photoemission spectra (XPS) measurements show that Pd 4*d* and Se 4*p* core level peaks are located at 336.8 eV and 54.8 eV, respectively, indicating a clean mono-valence phase and the valence state of Pd as 1+. (Supplementary Figure S1) This is unexpected because, in usual square planer environments, the valence state of Pd is known to be 2+, *i.e.* $Pd^{2+}$ ($d^8$) with $(Se_2)^{2-}$ neighbors, like the valence state of cation in typical pyrite-type transition metal dichalcogenides[12,20,21].

Figures 1c, 1d, and 1e show the electrical conductivity ($\sigma$), carrier concentration ($p$), and Hall mobility ($\mu_H$), respectively, as a function of temperature ($T$) from 80 to 313 K. As shown in Figure 1c, $\sigma$ shows exponential drop as $T$ decreases exhibiting thermally activated behavior (linear relation between $ln\ \sigma$ and $1/T$), which denotes the semiconducting transport nature with a finite $E_g$[22]. The valence photoemission spectra also indicate the semiconducting

ground state of $PdSe_2$. (Supplementary Figure S2) The positive sign in the Hall-effect measurement reveals the hole character of the primary charge. $T$ dependence of $p$ well obeys the Arrhenius relation of $p \sim exp\ (-\Delta E/2kT)$, where $\Delta E$ is the activation energy, and $k$ is Boltzmann constant, as shown from Figure 1d. By fitting $ln\ p$ versus $1/T$ plot, two distinct $\Delta E$ can be extracted as 0.89 eV and 0.07 eV at high (> 295 K) and low $T$ region (< 260 K), respectively. Considering the energetic scale of measured $T$ compared to reported $E_g$ values[15,22], the high and low $T$ region should be referred to as an intrinsic and ionization range, respectively, showing that $E_g$ of 0.89 eV and donor ionization energy ($E_d$) of 0.07 eV[22]. $T$ dependence of $\mu_H$ can be plotted from the relation of $\sigma = pq\mu_H$. ($q$ is an elementary charge) The $\mu_H$ decreases upon increasing of $T$, confirming that the phonon-related scattering is dominant over impurity one down to 80 K[22]. It suggests the clean sample quality of our $PdSe_2$ single crystal relatively free from any structural and/or chemical defects in accordance with the XRD and Raman results. (Figure 1a and 1b)

To investigate the electronic structure of $PdSe_2$, we systematically performed ARPES measurement as presented in Figure 2. Two Pd atoms exist in the unit cell per layer (red box) as shown in the top view of the crystal structure. (Figure 2a) The constant energy surfaces (CESs) at Γ-slice of $E_b$ = 0.8 eV and 1.6 eV are displayed with the corresponding DFT results in Figures 2b and 2c, respectively. (Complementary Brillouin zone (BZ) is indicated with the same color of Figure 2a.) As can be seen, the measured and calculated electronic structures with unfolding show very good agreements. Note that, in fact, the periodicity of ARPES spectral weight is not matching with the periodicity of the DFT band structures. (See Figures 2e and 2f) Because all Pd atoms, as well as Se-Se dimers in a unit cell, have similar orbital character, which is indeed shown from the shapes of the corresponding wave function (the inset of Figure 3e), the electronic structure mainly follows a reduced cell (blue box) which contains a single Pd atom and a Se-Se dimer instead of the orthorhombic unit cell (red box). To better account for this, we have unfolded the band structure, which successfully describes the experimental spectral weight as well as the CESs as shown in Figures 2b and 2c. (Supplementary Figure S3) The spectral weight of a hole band is clear at Γ point and dimly exists at Γ" point but vanishes at Γ' point. The weak spectral weight of Γ" stems from the inevitable $k_z$ broadening because Γ (Γ") is Γ (Z) while Γ' becomes X (or Y) in the unfolded BZ. This evidently shows that the character of the top valence band is barely influenced by the cell-doubling potential induced by the zigzag stacking of the Se-Se dimers.

The calculated band structure and the measured ARPES spectra are overlapped in Figures 2d-2f demonstrating the excellent agreement between them. Not only the binding energy of the valence bands but also the detailed band dispersions and bandwidths are well-reproduced from the DFT calculation. While the top valence band follows the unfolded electronic structure, shadow bands are observed below $E_b$ = 1 eV. These bands are originated from the cell-doubling potential, mainly due to the small deviation of $Se_2$ orbital from $p_z$-like one, which breaks the translation symmetry of reduced cell. (See the inset of Figure 3e.) The measured ARPES spectra display additional spectral features which can be attributed to the broadening of $k_z$ space. (Supplementary Figure S4)

The most interesting feature from the band structure in Figure 2d is the highly dispersive band along Γ-Z, which is even more dispersive than those along the in-plane directions. This is unprecedented considering the 2D nature of the $PdSe_2$ system including large interlayer space. In spite of some examples which exhibit the anionic bonding-induced dispersive bands along $k_z$ direction[23,24], $PdSe_2$ is not the case because the anionic bonding of Se-Se dimers is confined in the layer, and the interlayer Se-Se distance (3.75 Å) is much longer than the intralayer Se-Se distance (2.36 Å).

To have a more insight of the atypical $k_z$ dispersive band of $PdSe_2$, we further investigates the details of the spectra. The CES ($E_b$ = 1.2 eV) of $k_y$-$k_z$ plane was obtained by accumulating the photon energy dependent ARPES spectra at $k_x = 0$ cut (Figure 3a.). The closed CESs at the 6th and 8th Γ points are clearly observed, but the spectral weights at the 5th and 7th Γ points are relatively weak and the electronic structure does not follow the periodicity of folded BZ (red box). As we have pointed out, a unit cell contains two $PdSe_2$ layers and the electronic structure obeys the unfolded BZ (blue box). The unfolded CES map from DFT calculation in Figure 3b well-matched with the experiments in Figure 3a. One can observe shadow spectral weight in the 5th and 7th Γ points of both experimental and computational CES maps. The ARPES spectra in Figure 3c clearly demonstrates that the valence band along Γ-Z is highly dispersive, the electronic structure follows 2Γ periodicity, and the spectral weights below $E_b$ = 2 eV are prominent at 5th and 7th Γ region. Furthermore, the ARPES data exhibits sinusoidal-wave-like $k_z$ band with about 2.5 eV bandwidth. This is a rare case with extreme out-of-plane dispersion in the TMD systems, which is generally considered as 2D materials.

Figure 3d shows the projected density-of-state (PDOS) plots of Pd 4*d* and Se 4*p* orbitals with respect to the crystal axes. We see that the conduction band minimum (CBM) mainly consists of the Pd $d_{x^2-y^2} + d_{xy}$, and Se $p_x + p_y$ / $p_z$ while the valence band maximum (VBM) dominantly consists of Pd $4d_{3z^2-r^2}$. This gives important information on the local electronic structures: First, Se atoms form a dimeric molecular orbital $(Se_2)^-$ which possesses large portion of unoccupied antibonding states above Fermi level. Because the Se-Se molecule are bonded in the form of $(Se_2)^-$ instead of $(Se_2)^{2-}$, the valence state of Pd consequently becomes $Pd^+$ ($4d^9$) instead of $Pd^{2+}$ ($4d^8$); Second, with the valence state of $Pd^+$ $(Se_2)^-$, the unoccupied Pd 4*d* states has in-plane character ($d_{x^2-y^2}$ and $d_{xy}$ indicated by the blue arrow) while the VBM top has out-of-plane character ($d_{3z^2-r^2}$ indicated by the red arrow). Since the $Pd^+$ ion is located in a square planar symmetry, the ligand field splitting pushes up the in-plane orbital states above the chemical potential. (Supplementary Figure S5 for detailed analysis of the crystal electric field levels.) This strongly anisotropic electronic character from PDOS is experimentally verified by X-ray linear dichroism (XLD) measurement as shown in Figure 3e. In the Pd $M_{3,2}$-edge XLD spectra, the white line signal for in-plane polarization (E//ab) is compared to that of out-of-plane one (E//c), demonstrating the in-plane polarized 4*d* states for CBM as predicted from PDOS plot (blue arrows). Furthermore, the wave function plot in the inset of Figure 3e shows that the VBM is dominated by Pd $4d_{3z^2-r^2}$ whose orbital directionality prefers crystallographic *c*-axis, which results in exceptionally dispersive valence band along Γ−Z direction. Note that slightly tilted $p_z$ orbitals of Se form antibonding orbital within the Se dimer, bridging the Pd $d_{3z^2-r^2}$ orbitals along *c*-axis with antibonding coupling. The strong directional character of Pd and Se dimers is responsible to the strong interlayer bonding along *c*-axis as well as the large Γ-Z dispersion.

This unique orbital character around the Fermi level directly affects the layer-dependent behavior in 2D $PdSe_2$[12-17]. For the detailed investigation on the correlation between the electronic structure and the number of the layers, we performed systematic DFT calculations from monolayer to bulk systems as shown in Figure 4. The PDOS plots in Figure 4a exhibit the strong Pd $d_{3z^2-r^2}$ and Se $p_z$ character of the VBM for all layers. For the monolayer case, enhanced DOS just below the Fermi energy comes from the flat bands. (Refer to the band structure plot in Figure 4d.) As the number of layers increases, the number of interlayer hopping channels increases while the intralayer ones remain almost inert, which results in the systematic evolution of the $d_{3z^2-r^2}$-type hole bands: One can observe the splitting of a Pd $d_{3z^2-r^2}$ states in monolayer case into two bonding-antibonding states in bilayer, again into four states in 4 layers in PDOS (denoted with short black ticks in Figure 4a), which eventually form a dispersive bands for bulk case. This is schematically described in Figure 4b. As the band gap is formed between the $d_{3z^2-r^2}$-type hole states and unoccupied electron states, the layer-dependent gap evolution can be explained as in the schematic Figure 4b. Also, the calculated band gap upon layer numbers in Figure 4c is well-fitted with simple interlayer-hopping model[9]. The orbital-projected band structures in Figure 4d show the VBM mostly with Pd $d_{3z^2-r^2}$ character for all layer ranges, again indicating the strongly directional bonding behavior. This exceptionally directional Pd orbital character, which is associated with the unusual $4d^9$ valence state, is a main player for the unique layer-dependent properties found in puckered 2D $PdSe_2$ as distinct from typical TMD systems.

## Conclusions

We demonstrate the origin of strong interlayer interaction in puckered 2D $PdSe_2$ based on rigorous experimental and theoretical analysis on single crystalline $PdSe_2$ sample. From electrical transport and photoemission measurements, electrical ground state can be clarified as a semiconductor with a finite $E_g$. ARPES spectra clearly show highly dispersive valence band structures both for in-plane and out-of-plane direction, strongly evidencing that the degree of interlayer coupling is comparable to that along intralayer direction even under a large distanced interlayer spacing. We show that the strong interlayer coupling behavior in $PdSe_2$ is a consequence of the unique directional orbitals of $Pd^+$ and $(Se_2)^-$ dimers in conjunction with the square planar local environment. This exceptional coupling behavior results in the sensitive $E_g$ tunability with layer numbers, which we demonstrate with the systematic DFT approaches. We believe that these findings shed light on the understanding of underlying physics and chemistry associated to the novel interlayer coupling which is beyond simple vdW picture in 2D materials.

## Methods

**Sample synthesis.** Single crystalline 2D $PdSe_2$ was synthesized by the melt-solidification method[25]. After mixing of Pd and Se powders in the ratio of 1:2, we pressed them into a pellet form, which was subsequently sealed in an evacuated silica tube. The vacuum sealed sample was annealed at 850 $^oC$ for 50 hrs, then slowly cooled in the electric furnace down to 450 $^oC$ at a cooling rate of 1 $^oC$/hr. After reaching 450 $^oC$, the sample was cooled to room temperature by turning off the furnace. Finally, centimeter-sized $PdSe_2$ single crystal with a well cleaved in-plane surface was obtained as shown in the inset photo of Figure 1a.

**Structural, optical, and electrical characterization.** The crystalline orientation of the synthesized sample was characterized by XRD using a PANalytical diffractometer model Empyrean (Cu $K_\alpha$). The Raman spectrum was measured by Raman spectrometer (RAMANtouch, nanophoton) with 532 nm wavelength excitation. For the electrical characterization, we applied Ag paste at four corner edges of single crystal plate ($1\times1\times0.2$ $cm^3$), resulting in a van der Pauw geometry contact. The electrical transport properties under a various $T$ (from 80 to 313 K) were investigated using a cryogen-free cryostat equipped with a 1.2 T electro-magnet. (Co-designed by I.V SOLUTION and Sungwoo Instruments Inc.) The $\sigma$ and Hall voltage were measured using the standard van der Pauw technique by utilizing high-precision current source, nanovolt meter, and Hall-effect card. (6220, 2182A, and 3756, Keithley, respectively)

**Spectroscopic analysis.** Angle resolved photoemission and soft x-ray absorption measurements were performed in the 4A1 μ-ARPES and 2A MS beamline of the Pohang Light Source, Korea. The photoemission spectra were collected by using VG-Scienta R4000 analyzer. A clean surface was prepared by the top-post cleaving method at 300 K and the vacuum condition was better than $10^{-10}$ Torr. The $k_z$ dependence spectra were measured by covering a wide range of photon energy (54 ~ 190 eV). The inner potential $V_0$ = 10.5 eV was employed to obtain the crystal momentum $k_z = 0.512\sqrt{E_{kin}\sin^2\theta + V_0}$ (Å$^{-1}$) along the crystal $c$-axis. The presented ARPES spectra of Γ and Z cut plane were measured by $E_{ph}$ = 86 eV and 72 eV, respectively. The core level spectra were measured with $E_{ph}$ = 600 eV. The X-ray absorption spectra were measured using total electron yield in which the beam and sample drain currents were simultaneously measured to normalize the beam intensity. The grazing incident angle of photon was 70° and EPU72 enables us to measure E//c and E ⊥ c spectra without changing the experimental geometry.

**Density functional theory calculations.** Electronic structure calculations were mostly performed employing the projector-augmented-wave method implemented in the Vienna *ab* initio simulation package (VASP)[26,27]. Energy cutoff of 350 eV is used for the plane-wave expansion and 12 × 12 × 8 Monkhorst-Pack $k$-mesh sampling for bulk case[28]. Various functionals were tested to reproduce the experimental lattice parameters and band gap. (See Supplementary Figure S5 and related description.) Among them, we adopted the SCAN semilocal functional in the main manuscript[29,30]. We have fully optimized the lattice parameters as well as the internal positions with force criteria of 0.01 eV/Å. The unfolded electronic structures were presented with PBE functional with full-potential local-orbital minimum basis (FPLO) code[31].

**Acknowledgements**

This work was supported by the National Research Foundation of Korea (NRF) grant funded by the Korea government (MSIT) (No. NRF-2021R1A4A1031920, No. 2016K1A4A4A01922028, and No. 2019R1F1A1052026). K.K. was supported by the NRF grant (No. 2016R1D1A1B02008461), and the internal R&D program at KAERI (No. 524460-21). B.K acknowledges support by NRF Grant No. 2021R1C1C1007017 and KISTI supercomputing Center (Projects No. KSC-2020-CRE-0279). K.-T.K. was supported by the internal R&D program at KBSI (No. C140110).

**Author contributions**

J.H.R and K.L synthesized $PdSe_2$ single crystals and performed XRD, Raman, and transport property measurements. J.-G.K., B.-G.P. and K.-T.K. performed ARPES measurements and J.-G.K., Y.K. and K.-T.K. contributed to XAS measurements. B.K., K.K. and S.K. conducted the DFT calculations and analyses. B.K., K.-T.K., and K.L. directed the study and wrote the manuscript with the discussions and contributions of K.K and S.K.

**Figure Legends**

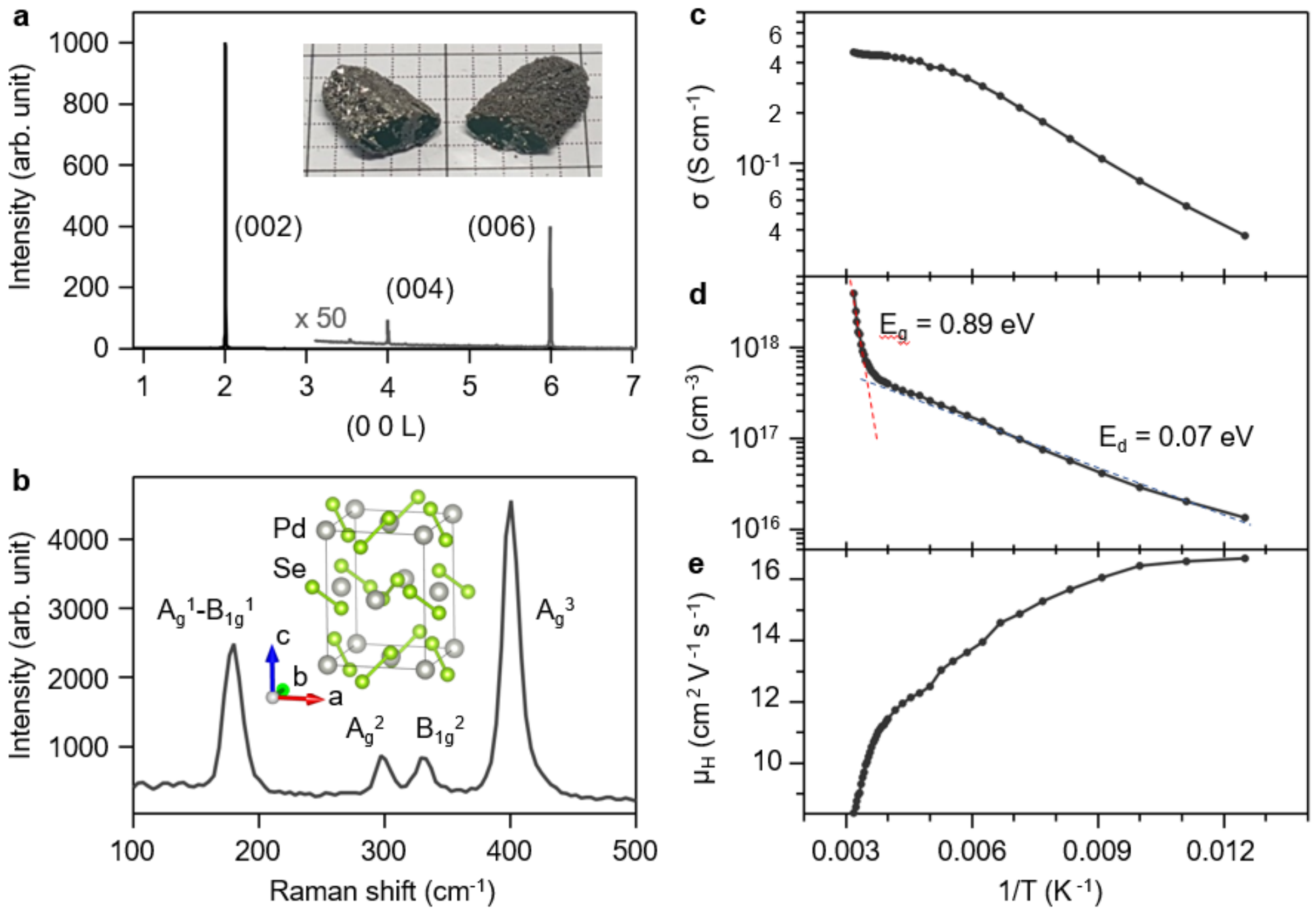

**Figure 1. a.** XRD pattern **b.** Raman spectrum on the cleaved surface of single crystalline $PdSe_2$ as shown like the inset photo of Figure 1a. Schematic of crystal structure is illustrated in the inset of Figure 1b. *T* dependence of **c.** $\sigma$, **d.** $p$, and **e.** $\mu_H$ of single crystalline $PdSe_2$.

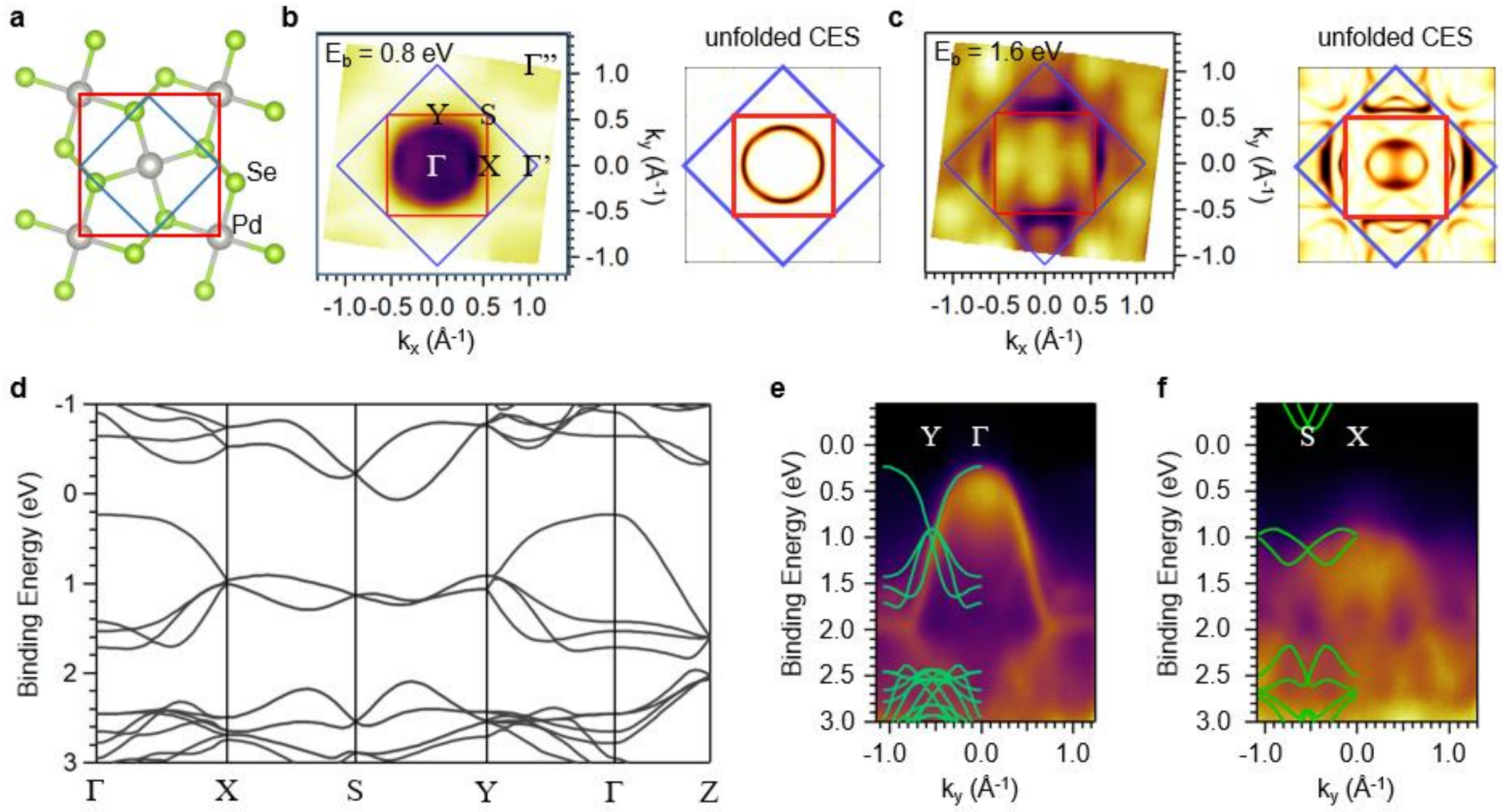

**Figure 2. a.** Top view of crystal structure of $PdSe_2$. Red (blue) box indicates crystallographic (reduced) unit cell. CESs for **b.** $E_b$ = 0.8 eV and **c.** 1.6 eV are displayed with unfolded DFT result. Red and blue boxes indicate the crystallographic and unfolded Brillouin zones, respectively. **d.** Band dispersion from SCAN calculation along Γ plane and Γ-Z direction. ARPES spectra of **e.** Y-Γ-Y and **f.** S-X-S cuts are displayed with calculated bands. (Green lines)

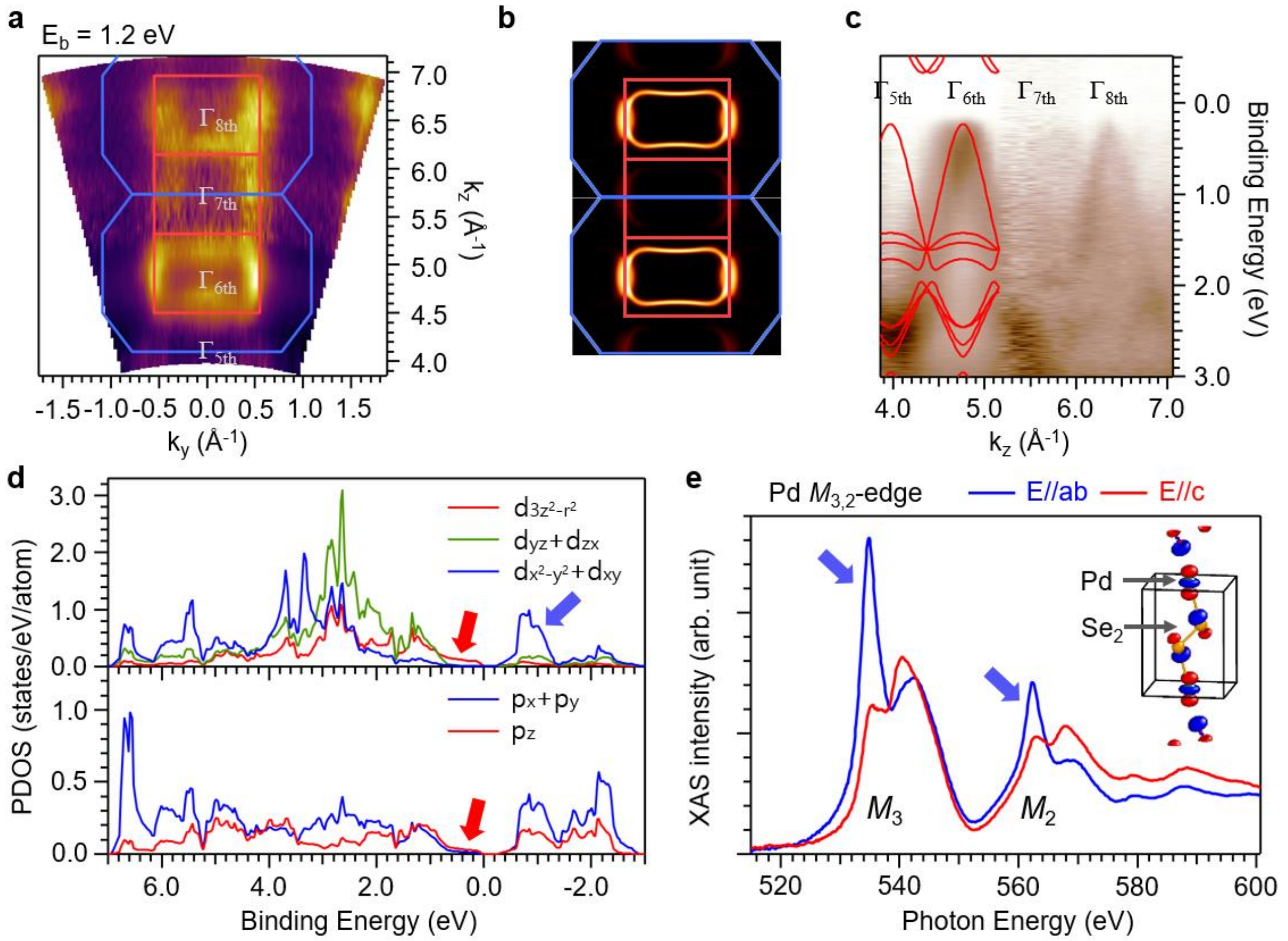

**Figure 3. a.** The $k_z$ CES map ($E_b$ = 1.2 eV) of Γ-Y cut obtained from photon energy dependent measurement. **b.** The equivalent CES map from DFT calculation. Red (blue) box indicates the crystallographic (unfolded) BZ. **c.** ARPES spectra along Γ-Z are displayed with calculated bands. **d.** PDOS plots of Pd 4*d* and Se 4*p* with respect to the crystal axes. Out of plane (in-plane) orbital characters of VBM (CBM) are indicated by red (blue) arrows, respectively. **e.** XAS-LD spectra of Pd $M_{3,2}$-edge. In-plane polarized unoccupied orbital states result in the large absorption white line as indicated by blue arrows. The wave function of VBM from DFT visualizes the orbital character of the $k_z$ dispersive band.

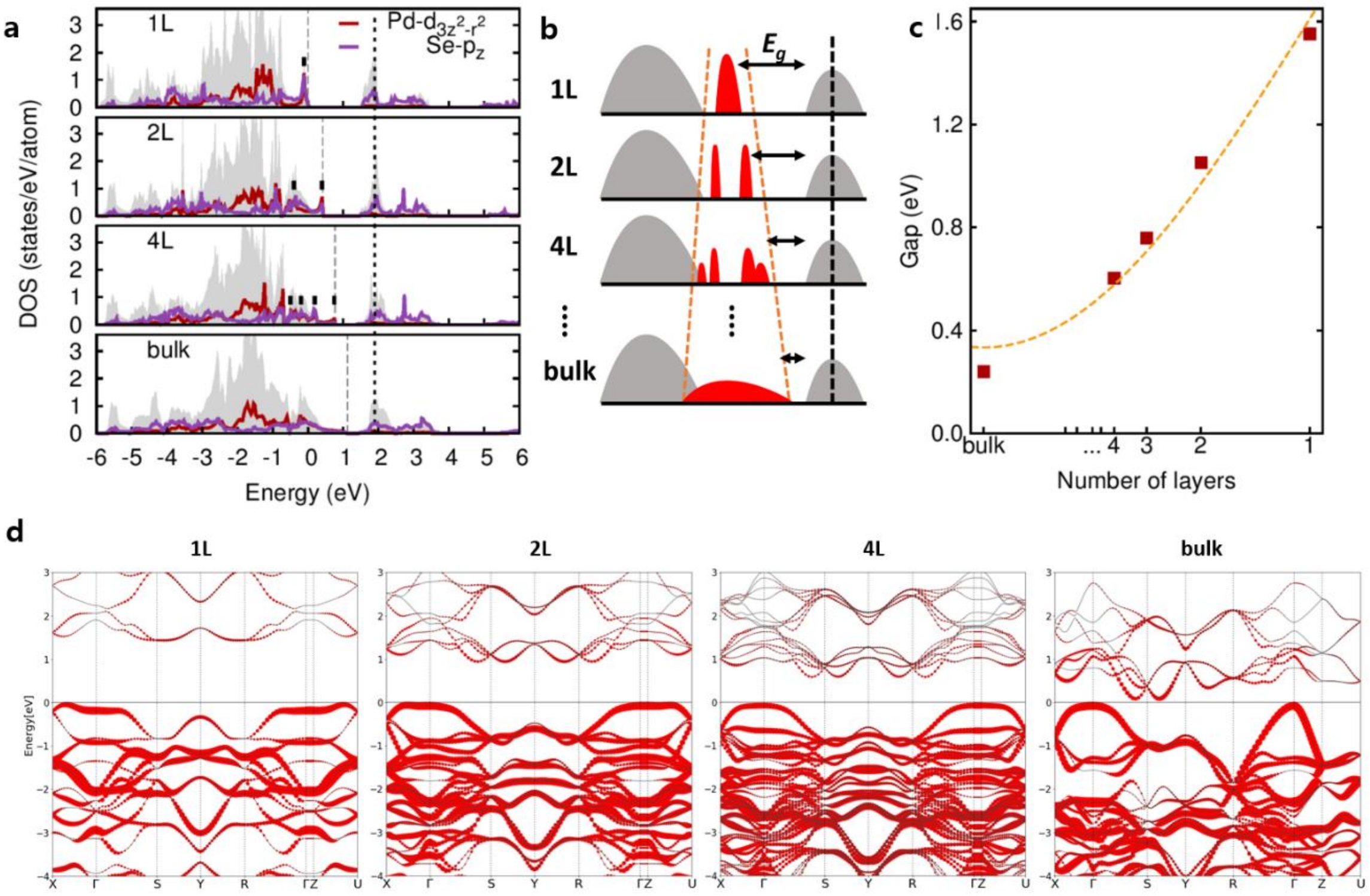

**Figure 4.** **a**. The layer-dependent evolution of the PDOSs. The position of VBM for each case is denoted with gray dashed line with aligning CBM. The black ticks indicate the splitting of $d_{3z^2-r^2}$ depending on the number of layers. **b.** The schematic diagram explaining the electronic structure upon layer numbers. **c.** Calculated energy gap as a function of the layer numbers. The yellow dashed line is fitted curve with simple interlayer-hopping model.[9] **d.** The band structure obtained from DFT calculations. The contribution of $4d_{3z^2-r^2}$ states is represented with the size of the red circles.

**Supplementary Information for**

# Direct observation of orbital driven strong interlayer coupling in puckered two-dimensional $PdSe_2$

Jung Hyun Ryu[1†], Jeong-Gyu Kim[2†], Bongjae Kim[1†], Kyoo Kim[3], Sooran Kim[4], Byeong-Gyu Park[5], Younghak Kim[5], Kyung-Tae Ko[6]*, and Kimoon Lee[1]*

[1]Department of Physics, Kunsan National University, Gunsan 54150, Republic of Korea.

[2]Max Planck POSTECH/Hsinchu Center for Complex Phase Materials and Department of Physics, Pohang University of Science and Technology (POSTECH), Pohang 37673, Republic of Korea.

[3]Korea Atomic Energy Research Institute (KAERI), Daejeon 34057, Republic of Korea.

[4]Department of Physics Education, Kyoungpook National University, Daegu 41566, Republic of Korea.

[5]Pohang Accelerator Laboratory, Pohang University of Science and Technology (POSTECH), Pohang 37673, Republic of Korea.

[6]Korea Basic Science Institute (KBSI), Daejeon 34133, Republic of Korea.

*e-mail : kkt0706@kbsi.re.kr, kimoon.lee@kunsan.ac.kr

[†]These authors are equally contributed.

## S1. Core level photoemission

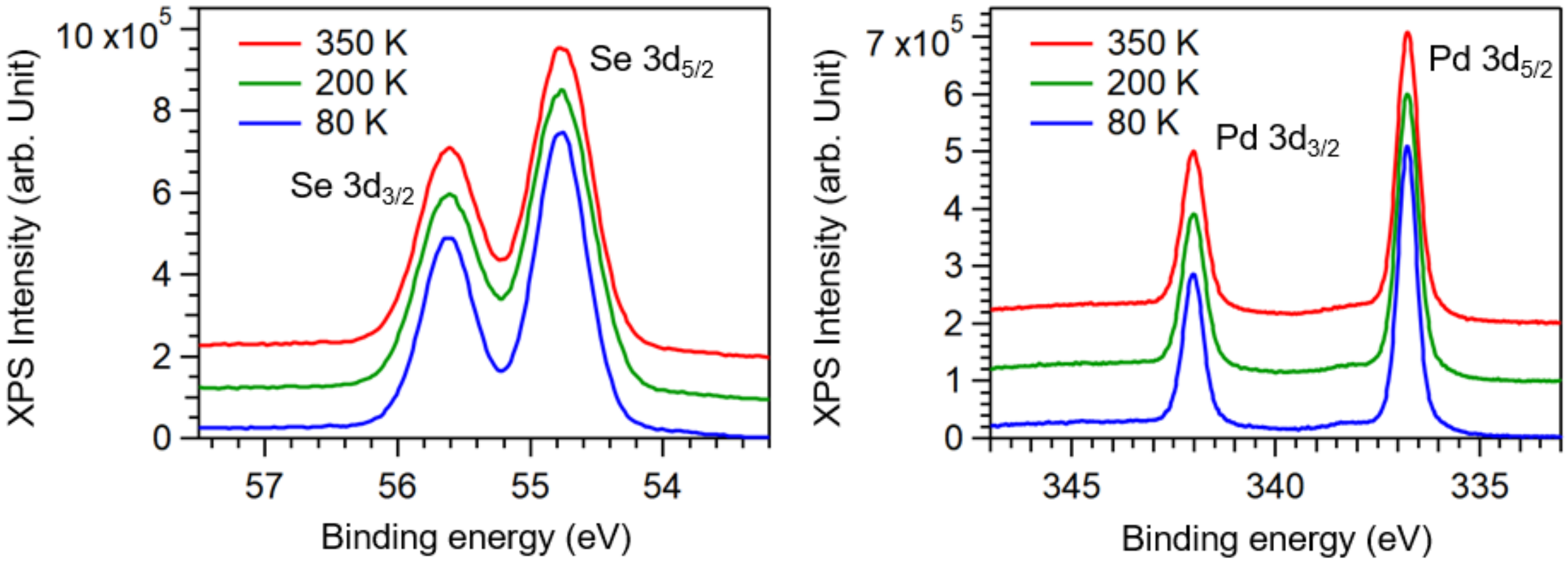

**Figure S1.** Se 3*d* and Pd 3*d* core-level photoemission spectra are measured at 4A1 beamline using $E_{ph}$ = 600 eV. The binding energy were calibrated using Au 4*f* spectra. One can clearly identify that the Se and Pd are monovalent. The symmetric profiles of peaks manifests the insulating nature of the system. $E_b$ (Se $3d_{5/2}$) = 54.8 eV and $E_b$ (Pd $3d_{5/2}$) = 336.8 eV show the covalent core-hole screening feature in $PdSe_2$, implying the Pd valence is largely departed from the ionic $Pd^{2+}$ case such as $PdCl_2$ ($E_b$ = 337.8 eV).[S1]

## S2. Valence band photoemission

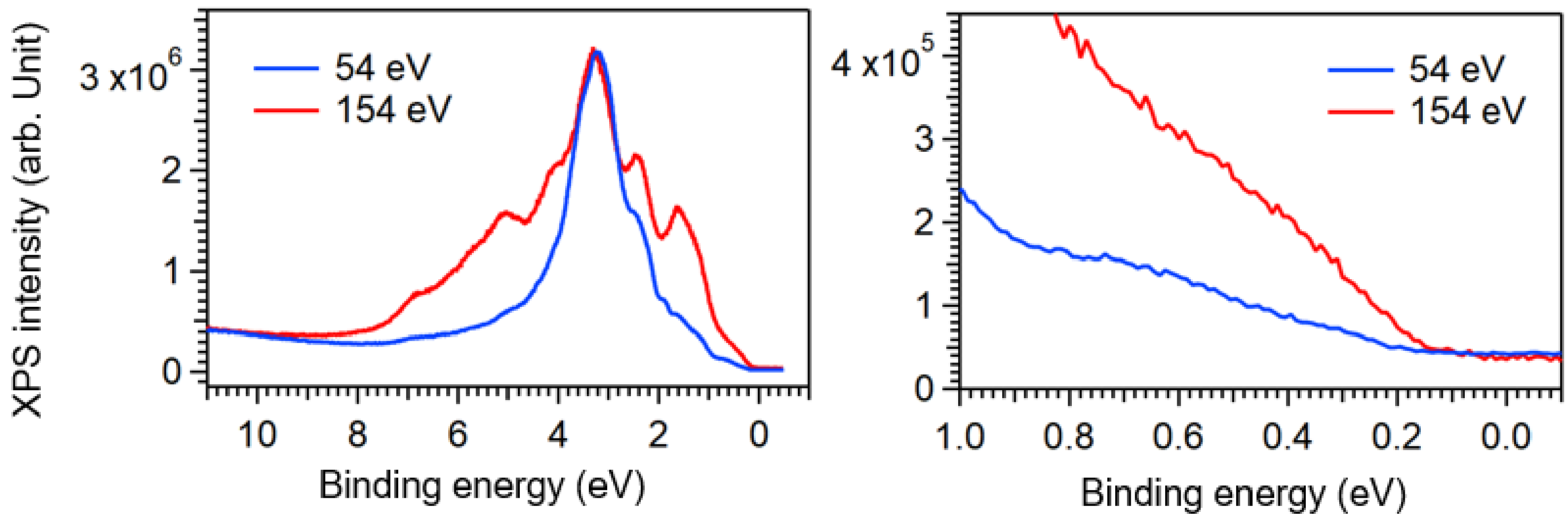

**Figure S2.** Valence photoemission spectra are measured with $E_{ph}$ = 54 and 154 eV. Results of full and narrow region are displayed in left and right panel, respectively. The blue line reflects the Pd 4d density of states (DOS) because the photoionization cross-section of Pd 4d at 54 eV is much larger than the one at 154 eV. This blue line spectra is well matched with calculated Pd 4d DOS in Figure 3d, where the Pd 4d weights mainly locate in the Eb = 2 ~ 4 eV window. The spectral weights vanish above $E_b$ = 0.2 eV at the valence band top, which is well-reproduced from the ARPES measurements and the DFT calculations. These results demonstrate that the insulating (semiconducting) nature of $PdSe_2$ containing a finite band gap.

## S3. Brillouin zone (BZ) for unfolding calculation

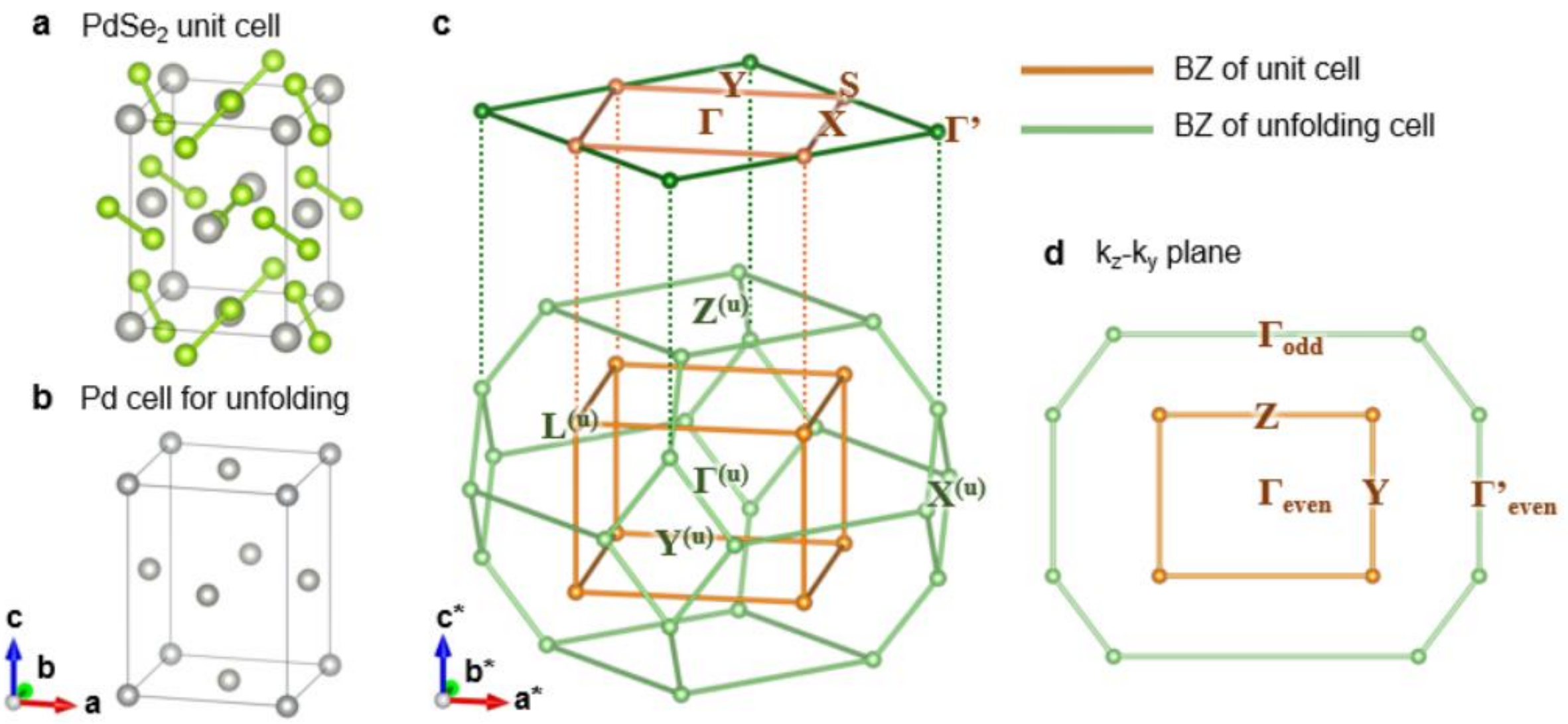

**Figure S3. a.** Crystallographic unit cell of $PdSe_2$. Zigzag structure of $Se_2$ dimer doubles the lattice periodicity. **b.** Pd only unit cell for the unfolding calculation. Because $3d_{z^2-r^2}$ orbital is insensitive to the perturbed potential of zigzag Se dimers, the band folding effects, *i.e.* spectral weight of shadow bands, are suppressed. **c.** BZs of crystallographic unit cell (orange) and face centered orthorhombic Pd cell (green) are displayed. The in-plane BZs using in Figure 2 are shown on top of 3D BZs diagram. **d.** The BZs diagram using Figure 3 is displayed. The crystal structure and the BZs are drown by using VESTA software.[S2]

**S4. Additional spectral features due to the broadening of $k_z$ space**

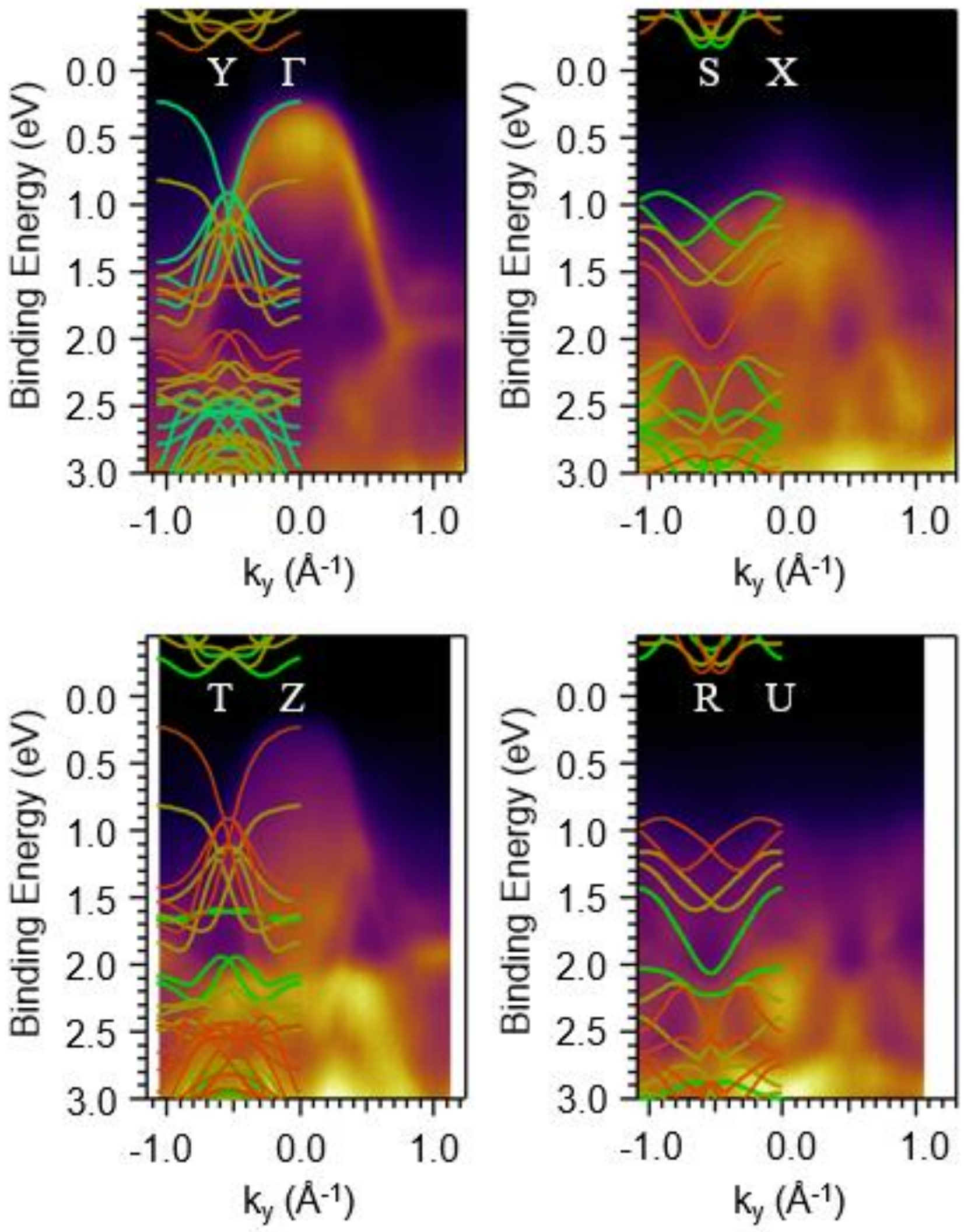

**Figure S4.** The ARPES spectra along high symmetry cuts on Γ and Z plane. Because the $k_z$ momentum is not conserved in the photoemission process, the $k_z$ broadening induces complicated ARPES spectra as shown in the figure. Green and red lines respectively represent the DFT calculated bands of target $k_z$ position and 0.5z* shifted position, and dark yellow line indicate 0.25z* shifted position. The measured spectral weight is well-matched not only with the green lines but also with the different colored lines. Despite the sizable $k_z$ broadening, we can resolve the $k_z$ dispersive electronic structure. The accumulated data clearly shows $k_z$ dispersive band as found in Figure 3. The nominal photon energy using the measurements was $E_{ph}$ = 86 eV (72 eV) for Γ (Z) plane.

## S5. Crystal field level

We have checked local crystal field splitting levels of Pd-*d* orbitals. The local *z*-axis of $PdSe_4$ square planar unit is slightly canted from the global axis as shown in Figure S5. The projected DOS well describes the crystal field level with the unoccupied $x^2$-$y^2$ hole at the conduction band minimum (CBM) and occupied $z^2$ states at the valence band maximum (VBM). As explained in the main manuscript, these levels can be schematically depicted with $d^9$ configurations rather than $d^8$ one, where the $\sigma$-type donor from Se is considered.

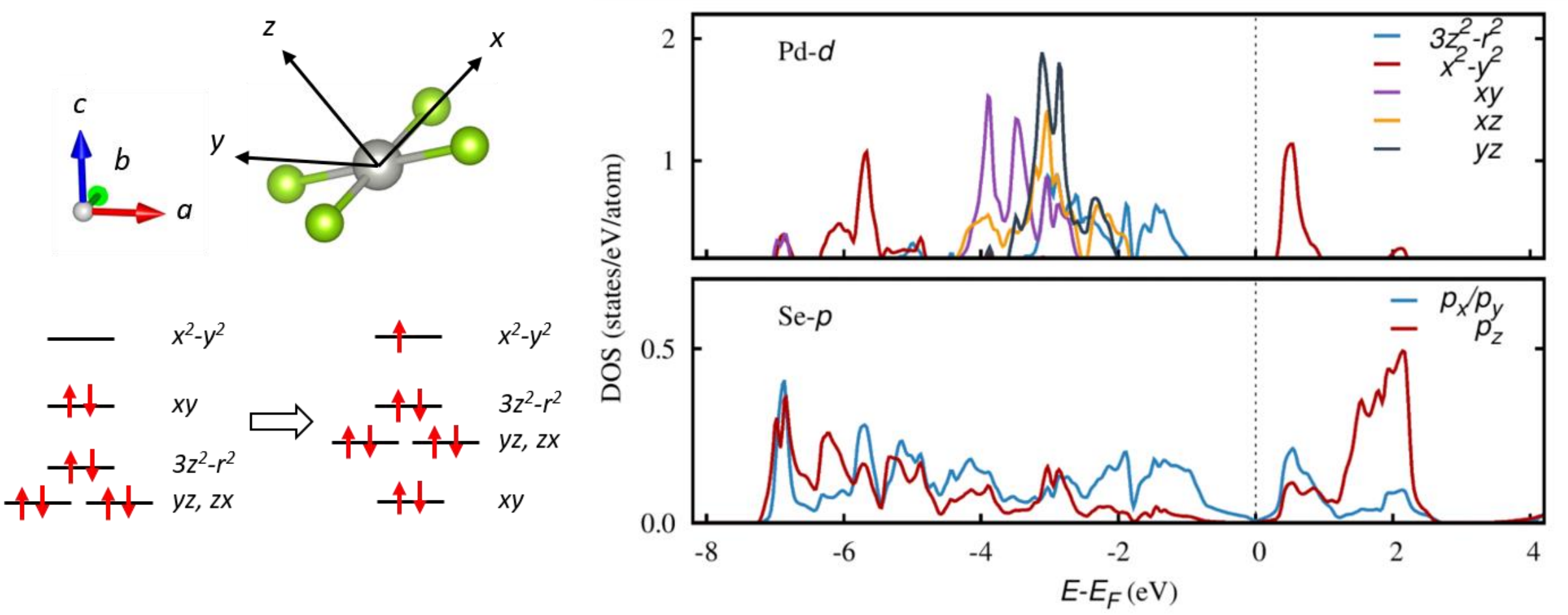

**Figure S5.** Local *d*-levels of Pd in the square planar environment. *x*,*y*, and *z*-axis and *a*, *b*, and *c*-axis denote the local and global axis respectively. The projected DOS clearly shows the well-split crystal field level of square planar unit with $d^9$ configurations

**S6. Comparison of various functionals**

To find the optimal functional for the system, we first systematically compared various functionals, which include local and semi local ones, with various van der Waals correction schemes. The results are summarized in Table S1. We found that in many cases, *c/a* ratio is severely underestimated without the van der Waals (vdW) interactions. By including the vdW contributions, either by correction or direct application of vdW density functional (vdW-DF), better description of the structures can be achieved. For example, TS21 and optPBE functionals generate reasonable value of *c/a* ratios and volumes but both cases predicted the system as metallic. In previous studies, optPBE vdW density functional has been readily employed, [S3-S5] which well-captures the *c/a* ratio, but this case cannot capture the gapped feature of the bulk $PdSe_2$ system. Here, we have chosen SCAN functional for most of the calculations, which reasonably reproduces not only the *c/a* ratio and volume of the system but also insulating electronic structures. Note that for the improved estimation of the band gap, advanced calculation such as GW methods are needed to include the long-range screening effects as well as the self-energy corrections, [S6] which is beyond the scope of our study.

Figure S6 shows the band structures of a few tested functionals, which reproduce the comparable *c*/*a* value with the experiment except PBE functional. The details and the exact positions of the bands are different but the overall band structures does not vary significantly as long as the structural parameters are similar to the experimental values. For example, the detailed energy positions of the bands for PBE band structure are deviated from others due to the incorrect estimation of the structural parameters. In the band unfolding calculation, we have used PBE functional with the experimental structural parameters. We have also tested the spin-orbit coupling (SOC) term in the calculation, and found that the effects are minimal.

**Table S1.** The comparison of various functionals for the description of *c/a* ratio and volume. The band gap values obtained for each relaxed structure are also shown.

| | | *c/a* | Volume ($V/V_0$) | Gap (eV) |
|---|---|---|---|---|
| **exp** | | 1.34 | 1.00 | 0.89 |
| **functional** | **LDA** | 1.01 | 0.89 | metal |
| | **PBE[S7]** | 1.46 | 1.14 | 0.42 |
| | **PBEsol[S8]** | 1.00 | 0.91 | metal |
| | **SCAN[S9,S10]** | 1.38 | 1.05 | 0.24 |
| **vdW-corrected** | **D3[S11]** | 1.00 | 0.93 | metal |
| | **D3-BJ[S12]** | 1.01 | 0.93 | metal |
| | **dDsC[S13]** | 1.02 | 0.93 | metal |
| | **MBD[S14]** | 1.28 | 1.01 | metal |
| | **TS20[S15]** | 1.28 | 1.03 | metal |
| | **TS21[S16]** | 1.29 | 1.03 | metal |
| **vdW-DF[S17]** | **optB96b** | 1.01 | 0.93 | metal |
| | **optB88** | 1.19 | 1.00 | metal |
| | **optPBE** | 1.34 | 1.08 | metal |
| | **revPBE** | 1.46 | 1.19 | 0.37 |
| | **SCAN-rVV10[S18]** | 1.34 | 1.02 | 0.05 |
| | **dvW-DF2[S19]** | 1.40 | 1.21 | 0.12 |

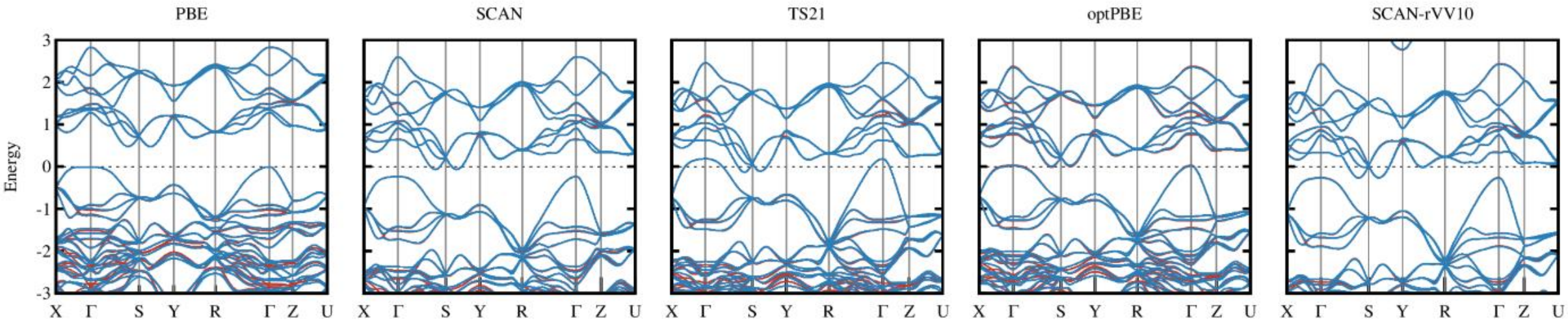

**Figure S6.** Band structures with fully relaxed structures for selected functionals. The blue and red curves are the band structures without and with SOC term, respectively.